\documentclass[twocolumn,prl,preprintnumbers,superscriptaddress,amsmath,amssymb,english]{revtex4}
\usepackage{amsmath,amssymb,amsfonts,mathrsfs}
\usepackage{amsthm}
\usepackage{CJK}
\usepackage{bm}
\usepackage{babel}
\usepackage{graphicx}
\allowdisplaybreaks
\usepackage{braket}
\usepackage[breaklinks]{hyperref}
\usepackage{color}
\usepackage{float}
\hypersetup{colorlinks=true, linkcolor=blue, citecolor=blue, filecolor=blue, urlcolor=blue}
\UseRawInputEncoding

\begin{document}

\title{Type-II Nodal Line Fermions in New $\mathbb{Z}_2$ Topological Semimetals \textit{A}$\mathbf{V_6Sb_6}$ (\textit{A}=K, Rb, and Cs) with  Kagome Bilayer}

\author{Y. Yang}
\affiliation{Hefei National Laboratory for Physical Sciences at Microscale and Department of Physics, and Key Laboratory of Strongly-Couple Quantum Matter Physics, Chinese Academy of Sciences£¬University of Science and Technology of China, Hefei, Anhui 230026, China}

\author{R. Wang}
\email[]{rcwang@cqu.edu.cn}
\affiliation{Institute for Structure and Function $\&$ Department of physics $\&$ Center for Quantum Materials and Devices, Chongqing University, Chongqing 400044, China}
\affiliation{Chongqing Key Laboratory for Strongly Coupled Physics, Chongqing 400044, China}

\author{M.-Z. Shi}
\affiliation{Hefei National Laboratory for Physical Sciences at Microscale and Department of Physics, and Key Laboratory of Strongly-Couple Quantum Matter Physics, Chinese Academy of Sciences£¬University of Science and Technology of China, Hefei, Anhui 230026, China}

\author{Z. Wang}
\affiliation{Hefei National Laboratory for Physical Sciences at Microscale and Department of Physics, and Key Laboratory of Strongly-Couple Quantum Matter Physics, Chinese Academy of Sciences£¬University of Science and Technology of China, Hefei, Anhui 230026, China}
\affiliation{CAS Center for Excellence in Superconducting Electronics (CENSE), Shanghai 200050, China}

\author{Z. Xiang}
\affiliation{Hefei National Laboratory for Physical Sciences at Microscale and Department of Physics, and Key Laboratory of Strongly-Couple Quantum Matter Physics, Chinese Academy of Sciences£¬University of Science and Technology of China, Hefei, Anhui 230026, China}

\author{X.-H. Chen}
%\email[]{chenxh@ustc.edu.cn}
\affiliation{Hefei National Laboratory for Physical Sciences at Microscale and Department of Physics, and Key Laboratory of Strongly-Couple Quantum Matter Physics, Chinese Academy of Sciences£¬University of Science and Technology of China, Hefei, Anhui 230026, China}
\affiliation{CAS Center for Excellence in Superconducting Electronics (CENSE), Shanghai 200050, China}
\affiliation{CAS Center for Excellence in Quantum Information and Quantum Physics, Hefei, Anhui 230026, China}
\affiliation{Collaborative Innovation Center of Advanced Microstructures, Nanjing University, Nanjing 210093, China}

\begin{abstract}
The recently discovered layered kagome metals \textit{A}$\mathrm{V_3Sb_5}$ (\textit{A}=K, Rb, and Cs) attract intensive interest due to their intertwining with superconductivity, charge-density-wave state, and nontrivial band topology. In this work, we show by first-principles calculations and symmetry arguments that unconventional type-II Dirac nodal line fermions close to the fermi level are present in another latest class of experimentally synthesized kagome compounds \textit{A}$\mathrm{V_6Sb_6}$ (\textit{A}=K, Rb, and Cs). These compounds possess a unique kagome $(\mathrm{V_3Sb})_2$   bilayer that dominates their electronic and topological properties, instead of the kagome V$_3$Sb monolayer in \textit{A}$\mathrm{V_3Sb_5}$. Crystal symmetry guarantees that the type-II Dirac nodal lines with quantized Berry phase lie in reflection-invariant planes of the Brillouin zone. We further reveal that the type-II Dirac nodal lines remain nearly intact in the presence of spin-orbital coupling and can be categorized as a $\mathbb{Z}_2$ classification. The findings establish \textit{A}$\mathrm{V_6Sb_6}$ as a class of new fascinating prototypes, which will extend the knowledge of interplay between unconventionally topological fermions and exotic quantum ordered states in kagome systems.
\end{abstract}


%Use showkeys class option if keyword %display desired

\maketitle
Interplay of symmetry, quantum ordered states, and nontrivial topology attracts intensive interest in condensed-matter physics.
Over the past decade, topological insulators and superconductors have been classified as a function of symmetry class \cite{RevModPhys.82.3045}, and various symmetry-protected nontrivial quasiparticles have been predicted or verified in gapless topological semimetals or metals \cite{PhysRevB.84.235126,PhysRevB.83.205101,Dirac2014,PhysRevX.5.011029,RevModPhys.90.015001}. Meanwhile, strict symmetries of free space are not necessarily preserved in crystalline solids, which means that unconventional topological quasiparticles without counterpart in standard model can emerge \cite{Bradlynaaf5037, ZhuPhysRevX.6.031003, NatureSoluyanov2015, PhysRevLett.117.066402, PhysRevLett.117.056805, Zhang2018}. As a representative example, two types of low-energy quasiparticles based on their band profiles have been identified in crystalline materials. The first type is the conventional type-I quasiparticles associated with a closed pointlike Fermi surface. The second type is the unconventional type-II quasiparticles, whose nodal points appear at the boundary of electron and hole pockets, leaving open trajectories \cite{NatureSoluyanov2015}. The low-energy excitations around a type-II nodal point do not satisfy the Lorentz symmetry and can lead to exotic quantum transport phenomena such as anisotropic chiral anomaly \cite{NatureSoluyanov2015, PhysRevLett.117.066402, PhysRevLett.117.056805}. Considering the diversity of crystal symmetries in crystalline solids, the realization of intertwining between unconventional topological quasiparticles and different quantum ordered states is intriguing.

The kagome lattice, possessing a two-dimensional network of tiled triangles and hexagons, is known as one of the most versatile models for studying various exotic quantum phenomena \cite{RevModPhys.89.025003}. Due to its unique lattice geometry, the kagome lattice hosts linear Dirac bands and flat bands, naturally generating cooperation of many-body correlated effects and nontrivial band topology, such as quantum spin liquid \cite{RevModPhys.89.025003,spin-liquid}, charge density waves (CDW) \cite{PhysRevB.80.113102}, superconductivity \cite{PhysRevB.79.214502}, as well as beyond \cite{Frustrated, PhysRevB.79.214502, PhysRevLett.97.147202,PhysRevB.87.115135, Mweyl-science1, Mweyl2, Mweyl3, 2020Quantum}. Recently, a new family of nonmagnetic kagome metals \textit{A}$\mathrm{V_3Sb_5}$ (\textit{A}=K, Rb, and Cs) has been discovered to host fascinating integration of multiple quantum phenomena \cite{PhysRevMaterials.3.094407, doi:10.1126/sciadv.abb6003, NatuecomChen, PhysRevX.11.031050, PhysRevX.11.031026, PhysRevLett.127.046401, PhysRevLett.125.247002, PhysRevMaterials.5.034801, 2021chiral, 2021Naturegao, 2021Natureze}. These materials possess a layered structure that crystallizes in space group P6/mmm (No. 191), with a prototypical V$_3$Sb plane composed of the V kagome sublattice and Sb hexagonal sublattice. Due to the presence of vanadium kagome net, immediate studies of collective ordered states including superconductivity and CDW have been intensively carried out in these \textit{A}$\mathrm{V_3Sb_5}$ kagome metals \cite{PhysRevX.11.031050, PhysRevX.11.031026, PhysRevLett.127.046401}. Moreover, the theoretical calculations and angle-resolved photoemission spectroscopy measurement identified that electronic band structures of \textit{A}$\mathrm{V_3Sb_5}$ feature nontrivial $\mathbb{Z}_2$ band topology \cite{PhysRevLett.125.247002, PhysRevMaterials.5.034801}. While these advancements of intimate correlations between superconductivity, charge order, and nontrivial band topology in \textit{A}$\mathrm{V_3Sb_5}$ have been very encouraging \cite{PhysRevX.11.031026, 2021chiral}, unconventional topological quasiparticles and their interplay with various quantum ordered states still remains largely unexplored. In addition, multiple crossing points and relative complicated Fermi surface in the electronic band structures of \textit{A}$\mathrm{V_3Sb_5}$ make their understanding by theoretical and experimental difficult. Therefore, it is highly desirable to explore kagome materials with unconventional and ideal topological band structures.

Very recently, we have discovered another class of layered kagome compounds \textit{A}$\mathrm{V_6Sb_6}$ with space group $R\bar{3}m$ (No. 166) \cite{A166arXiv}, which would satisfy the above criteria. These new kagome compounds \textit{A}$\mathrm{V_6Sb_6}$ have the same chemical compositions but different stoichiometric ratios with \textit{A}$\mathrm{V_3Sb_5}$. Compared with \textit{A}$\mathrm{V_3Sb_5}$, \textit{A}$\mathrm{V_6Sb_6}$ can be considered as interposing an additional V$_3$Sb plane; that is, a bilayer $(\mathrm{V_3Sb})_2$  stacking network is present in $\mathrm{AV_6Sb_6}$. Meanwhile, it is worth noting that the \textit{A}$\mathrm{V_6Sb_6}$ compounds as well as \textit{A}$\mathrm{V_3Sb_5}$ can be viewed as a unified \textit{A}-V-Sb family with a generic chemical formula \textit{A}$_{m-1}$Sb$_{2m}$(V$_3$Sb)$_n$ (i.e., \textit{A}$\mathrm{V_3Sb_5}$ with $m=2$, $n=1$ , and \textit{A}$\mathrm{V_6Sb_6}$ with $m=2$, $n=2$), exhibiting common features of this family. Importantly, the experimental results indicates that the \textit{A}$\mathrm{V_6Sb_6}$ compounds exhibit superconductivity under pressure \cite{A166arXiv}. In this work, based on first-principles calculations and symmetry analysis, we identify that the \textit{A}$\mathrm{V_6Sb_6}$ compounds are ideal $\mathbb{Z}_2$ topological semimetals and possess the symmetry-protected type-II Dirac nodal line fermions close to the Fermi level, implying that  \textit{A}$\mathrm{V_6Sb_6}$ could be expected as a promising platform for investigating exotic quantum phenomena with unconventional topological quasipariticles beyond standard model.

\begin{figure}
    \centering
    \includegraphics[scale=0.77]{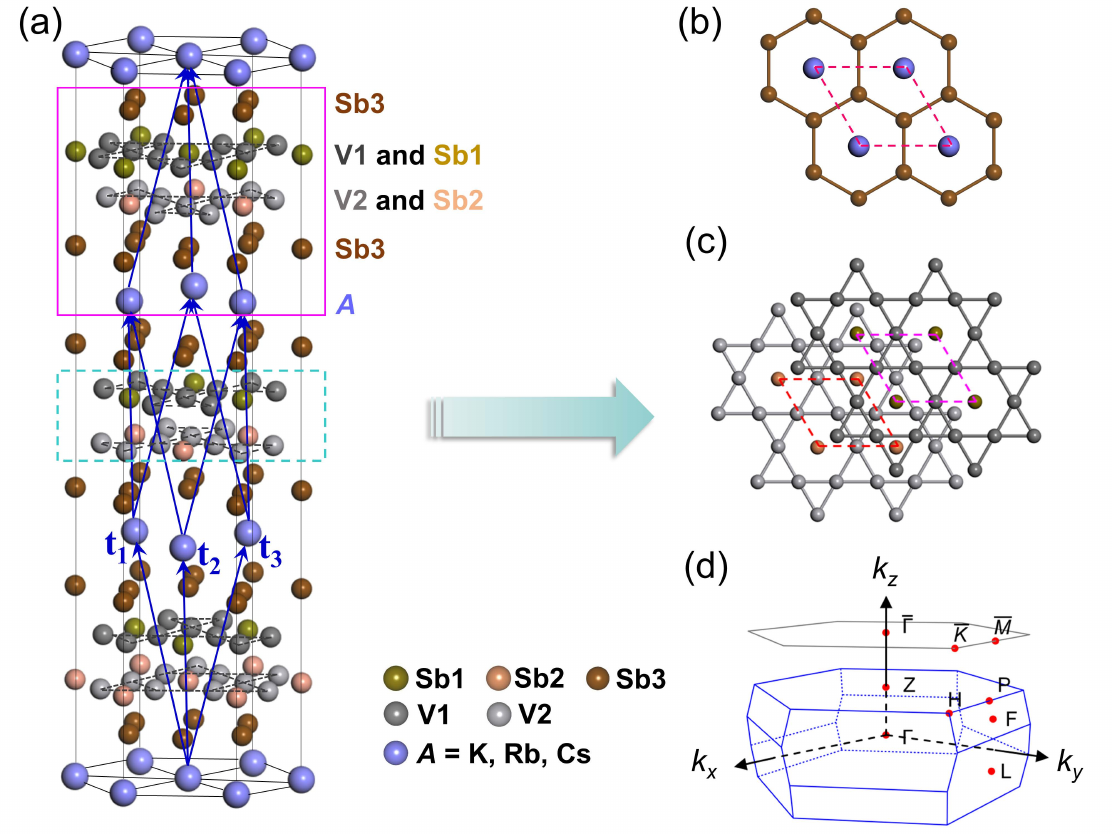}
    \caption{(a) The crystal structure of \textit{A}$\mathrm{V_6Sb_6}$ (\textit{A} = K, Rb, and Cs) with \textit{A}-Sb-V$_3$Sb-V$_3$Sb-Sb quintuple layers stacking along the crystallographic $c$-axis. $\mathbf{t}_{1,2,3}$ is the primitive lattice vectors of a rhombohedral structure with space group $R\bar{3}m$ (No. 166) \cite{SM}. (b) The graphitelike honeycomb sublattice of the Sb3 atom and hexagonal sublattice of \textit{A}-site alkali metal atom. (c) The V$_3$Sb bilayer network composed of two V1 and V2 kagome sublattices and two Sb1 and Sb2 hexagonal sublattices. (d) The rhombohedral BZ and the corresponding (111) surface BZ. High-symmetry points (solid red points) and a high-symmetry middle plane are marked.
    \label{FIG1}}
\end{figure}

To elucidate topological features of the novel kagome compounds \textit{A}$\mathrm{V_6Sb_6}$, we performed first-principles calculations based on the density functional theory \cite{PhysRev.140.A1133} as implemented in the Vienna \textit{ab initio} simulation package \cite{PhysRevB.54.11169}. The exchange-correlation functional was described by generalized gradient approximation with Perdew-Burke-Ernzerhof formalism \cite{PhysRevLett.77.3865}. The core-valence interactions were treated by projector augmented-wave potentials \cite{PhysRevB.59.1758}, and a plane-wave basis set with a kinetic-energy cutoff of 450 eV was used. The Brillouin zone (BZ) was sampled by a $12\times12\times12$ Monkhorst-Pack grid \cite{PhysRevB.13.5188}. % to simulate the rhombohedral structure.
The crystal structures were fully relaxed by minimizing forces of each atom smaller than $1.0\times10^{3}$ eV/{\AA}, and van der Waals interactions along the $c$-layer stacking direction was considered by the Crimme (DFT-D3) method \cite{D3}. The topological classification was confirmed by the $\mathbb{Z}_2$ invariants \cite{PhysRevLett.98.106803}, which are calculated from the parity eigenvalues at the time-reversal invariant momenta (TRIM) points using IRVSP package \cite{GAO2021107760}. We construct a Wannier tight-binding (TB) Hamiltonian based on maximally localized Wannier functions to reveal the topological features as implemented in the WANNIER90 package \cite{Mostofi2008}.

As illustrated in Fig. \ref{FIG1}(a), the presence of kagome V$_3$Sb bilayer net makes the \textit{A}$\mathrm{V_6Sb_6}$ structure hosting \textit{A}-Sb-V$_3$Sb-V$_3$Sb-Sb quintuple layers, which stack along the crystallographic $c$-axis. As a result, the \textit{A}$\mathrm{V_6Sb_6}$ compounds crystallize in a rhombohedral structure with space group $R\bar{3}m$ (No. 166). As shown in Fig. \ref{FIG1}(c), the bilayer V$_3$Sb stacking network contains two V1 and V2 kagome sublattices, which share two equivalent Sb1 and Sb2 hexagonal sublattices. The Sb3 atom respectively forms upper and lower graphitelike honeycomb lattices that encapsulate the V$_3$Sb kagome bilayer, and the hexagonal sublattice of A-site alkali metal atom spontaneously fills the space between the Sb3 sublattice [see Fig. \ref{FIG1}(b)]. As a result, the atom positions of \textit{A}$\mathrm{V_6Sb_6}$ structure are  Wyckoff $1a$ of A atom, $6h$ of V1 and V2 atoms, and $3c$ of Sb1, Sb2, and Sb3 atoms. The optimized lattice constants are in excellent agreement with the experimental values (see details in Supplemental Material (SM) \cite{SM}).
The rhombohedral BZ and the corresponding (111) surface BZ of \textit{A}$\mathrm{V_6Sb_6}$ are shown in Fig. \ref{FIG1}(d), with high-symmetry points indicated.  Since these three novel kagome materials share the similar features, we only present the results of $\mathrm{CsV_6Sb_6}$ in the main text. The results of $\mathrm{KV_6Sb_6}$  and $\mathrm{RbV_6Sb_6}$  are included in the SM \cite{SM}.

\begin{figure}
    \centering
    \includegraphics[scale=0.75]{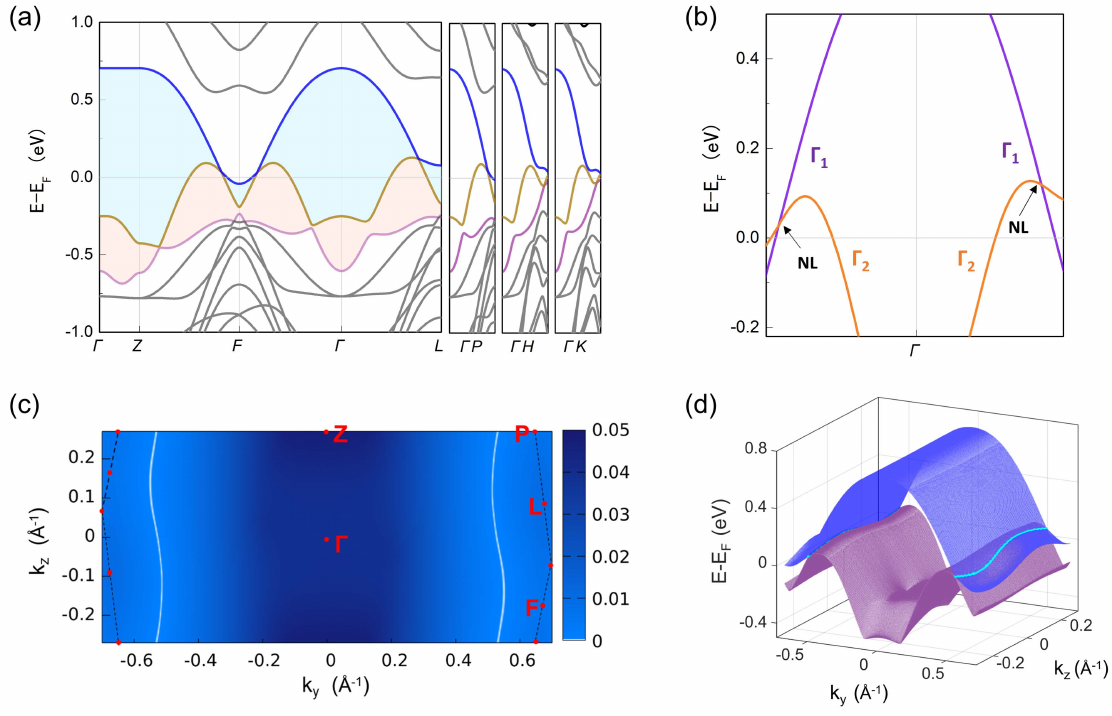}
    \caption{Electronic band structures of $\mathrm{CsV_6Sb_6}$ along high-symmetry lines in the absence of SOC. Three bands close to the Fermi level are highlighted in colors. (b) The enlarged view of band crossings along an arbitrary line in the $k_y$-$k_z$ plane, with irreducible representations of $C_s$ indicated. (c) Energy difference map between the conduction band and valence band  in the $k_y$-$k_z$ plane. (d) Two inverted bands that form two nodal lines, indicating that band profiles along the nodal line are tilted and thus belong to type-II.
    \label{FIG2}}
\end{figure}

The calculated electronic band structures of $\mathrm{CsV_6Sb_6}$ along high-symmetry lines are shown in Fig. \ref{FIG2}(a). The bands show that a set of linear band crossings are present along some specific high-symmetry lines. These high-symmetry lines are coplanar with the $k_z$ axis, forming a high-symmetry middle plane of the BZ [e.g., the $k_y$-$k_z$ plane with $k_x=0$ in Fig. \ref{FIG1}(d)]. Due to the existence of inversion ($\mathcal{I}$) and time-reversal ($\mathcal{T}$) symmetries, the crossing points in $\mathrm{CsV_6Sb_6}$ belong to Dirac points.  All the Dirac points are slightly above the Fermi level and their maximum energy at $\sim$0.09 eV occurs along $F$-$Z$, forming hole pockets. A deeper inspection reveals that the Dirac points can actually exist along an arbitrary line that connects the $\Gamma$/$Z$ point and the boundary point of the BZ in the $k_y$-$k_z$ plane [see Fig. \ref{FIG2}(b)]. We also plot an energy difference map between the valence band and conduction band in Fig. \ref{FIG2}(c). A zero energy difference denoted as the white feature indicates that there are two nodal lines threading the BZ in the $k_y$-$k_z$ plane. By checking the little group of the $k_y$-$k_z$ plane, we find that all Dirac points are with respect to the mirror-reflection symmetry $C_s$. Two crossing bands belong to opposite mirror eigenvalues $\Gamma_1$ (+1) and $\Gamma_2$ (-1). By the way, we should point out that the symmetry forbids the existence of Dirac points along $\Gamma$-$K$ though the conduction and valence bands seems to nearly touch [see Fig. \ref{FIG2}(a)]. A three-dimensional plot of two crossing bands in the $k_y$-$k_z$ plane with $k_x=0$ is present in Fig. \ref{FIG2}(d). This figure depicts that the two inverted  bands form two continuous Dirac nodal lines close to the Fermi level with tiny energy dispersion. Clearly, it can be found that band profiles near all Dirac points along the nodal lines are tilted, indicating that low-energy excitations in $\mathrm{CsV_6Sb_6}$ may belong to type-II Dirac nodal line fermions.

\begin{figure}
    \centering
    \includegraphics[scale=0.8]{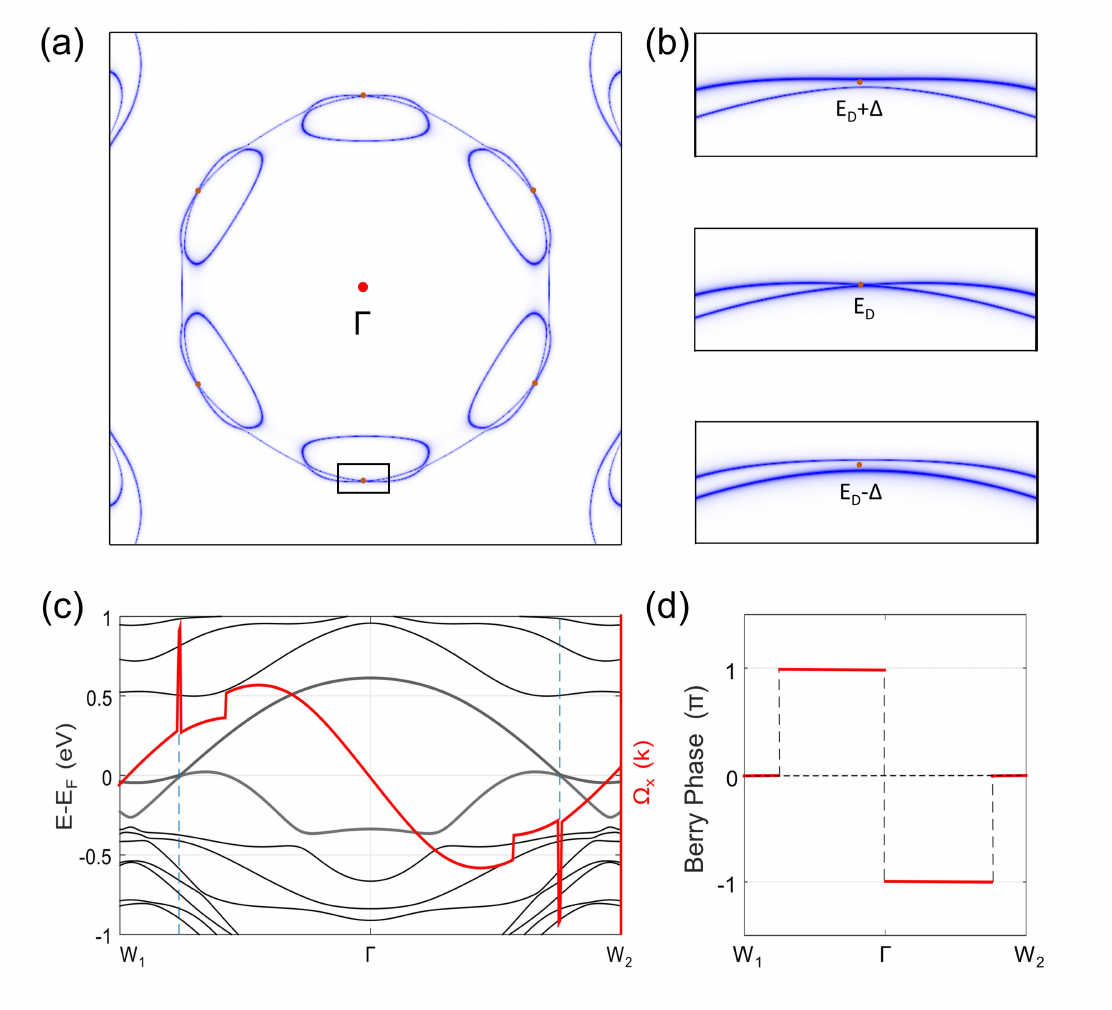}
    \caption{(a) The cross section of bulk isoenergy surface in the $k_x$-$k_y$ plane with $k_z=0$. (b) The enlarged view of isoenergy contours [marked by the black box in (a)] of $\Delta = 4$ meV below, at, and above the Dirac point $E_D$ are shown in the upper, middle, and lower panels, respectively. (c) The electronic band structures as well as Berry curvature along $W_1$-$\Gamma$-$W_2$ in the $k_y$-$k_z$ plane.  (d) A variation of Berry phase along $W_1$-$\Gamma$-$W_2$ in the $k_y$-$k_z$ plane. The points $W_1$ and $W_2$ are positioned at (0.0, -0.7, 0.0) and (0.0, 0.7, 0.0) {\AA}$^{-1}$, respectively.
    \label{FIG3}}
\end{figure}

We further examine isoenergy contours of Fermi surface to verify the presence of type-II Dirac nodal lines in $\mathrm{CsV_6Sb_6}$. The cross section of bulk isoenergy surface in the $k_x$-$k_y$ plane with $k_z=0$ is shown in Figs. \ref{FIG3}(a) and \ref{FIG3}(b). We can see that there are six symmetry-distributed Dirac points in the first BZ, which are generated from nodal lines across the $k_z=0$ plane, are present in the first BZ. When tuning the chemical potential of the isoenergy surface, hole and electron pockets only touch at the energy of Dirac points $E_D$ and exhibit open isoenergy contours [see Fig. \ref{FIG3}(b)]. Therefore, $\mathrm{CsV_6Sb_6}$ indeed hosts type-II Dirac nodal lines as its preservation of $\mathcal{I}$ and $\mathcal{T}$ symmetries. Considering the three-fold rotational symmetry $C_3$ in space group $R\bar{3}m$, there are three equivalent mirror planes and six type-II Dirac nodal lines in the BZ.

Next, we carry out a symmetry argument to understand the existence of six symmetry-protected type-II Dirac nodal lines in $\mathrm{CsV_6Sb_6}$. As shown in Fig. 1(b), the nonmagnetic space group $R\bar{3}m$ contains  three-fold rotational symmetry $\mathcal{C}_{3z}$ along the $z$ (or $c$) direction, three twofold rotational symmetries (i.e., one is the $\mathcal{C}_{2x}$ along the $x$ (or $a$) direction, and the other two are with respect to $\mathcal{C}_{3z}$), $\mathcal{I}$-symmetry, and $\mathcal{T}$-symmetry. In general, two inverted bands can be described by a two-band effective $\mathbf{k}\cdot \mathbf{p}$ Hamiltonian as
\begin{equation}\label{Hamiltonian}
\mathcal{H}(\mathbf{k})=f_0(\mathbf{k})\sigma_0+f_x(\mathbf{k})\sigma_x+f_y(\mathbf{k})\sigma_y+f_z(\mathbf{k})\sigma_z,
\end{equation}
where $f_0(\mathbf{k})$ is the kinetic term, $\sigma_0$ is the identity matrix, $\sigma_i$ ($i=x,y,z$) are three Pauli matrices, $f_i(\mathbf{k})$ are real functions, and wavevector $\mathbf{k}$ is relative to the inverted point. The kinetic term $f_0(\mathbf{k})$ only shifts Dirac points and thus can be ignored in Eq. (\ref{Hamiltonian}) in the following. The two inverted bands have the opposite eigenvalues of little group $C_s$, indicating that the $\mathcal{I}$-symmetry can be represented as $\mathcal{I}=\sigma_z$, which constrains the Hamiltonian as
\begin{equation}\label{I}
\mathcal{I}\mathcal{H}(\mathbf{k}) \mathcal{I}^{-1} =  \mathcal{H}(-\mathbf{k}).
\end{equation}
In the absence of SOC, the $\mathcal{T}$ symmetry can be represented by $\mathcal{T}=K$, where $K$ is the complex conjugate operator. The $\mathcal{T}$-symmetry indicates
\begin{equation}\label{T}
\mathcal{T}\mathcal{H}(\mathbf{k}) \mathcal{T}^{-1} =  \mathcal{H}(-\mathbf{k}).
\end{equation}
Based on Eqs. (\ref{I}) and (\ref{T}), the existence of $\mathcal{I}$ and $\mathcal{T}$ symmetries requires
\begin{equation}
f_x(\mathbf{q})\equiv 0, \ \ f_y(\mathbf{k})=-f_y(\mathbf{-k}), \ \  f_z(\mathbf{k})=f_z(\mathbf{-k}).
\end{equation}
The band crossings require $f_y(\mathbf{k}) =0$ and $f_z(\mathbf{k}) =0$, which indicate that there is a codimension one, and thus allowing nodal-lines in momentum space. The mirror-reflection symmetry $\mathcal{M}_x$ can be considered as the product of $\mathcal{I}$ and $\mathcal{C}_{2x}$ (i.e., $\mathcal{M}_x = \mathcal{I}\mathcal{C}_{2x}$). In the $k_y$-$k_z$ plane with $k_x=0$, the commutation relation $[\mathcal{M}_x, \mathcal{H}]=0$ indicates $f_y(0, k_y, k_z)\equiv 0$. Then, we can expand $f_z(0, k_y, k_z)$ and remain the lowest orders as
\begin{equation}\label{expreduced}
f_z(0, k_y, k_z)= A_z+B_z k_y^2+C_z k_z^2+D_z k_yk_z,
\end{equation}
in which the parameters $A_z$, $B_z$, $C_z$, and $D_z$ can be fitted from first-principles calculations. The condition of open nodal lines in the $k_y$-$k_z$ plane requires that $A_z<0$, $4B_zC_z<D_z^2$ and $B_z>0$  or $4B_zC_z>D_z^2$ and $B_z<0$. Similarly, the other two mirror planes can also possess two same type-II nodal lines due to $\mathcal{C}_{3z}$. It should be mentioned that the stability of type-II Dirac nodal lines in $\mathrm{CsV_6Sb_6}$ is topologically protected by the coexistence of $\mathcal{I}$ and $\mathcal{T}$ symmetries, and the additional rotational and mirror symmetries just force nodal lines to locate in the mirror planes of the BZ.

The presence of type-II Dirac nodal lines corresponds to a quantized Berry phase or winding number with $\pi$ mod $2\pi$, which is defined as \cite{PhysRevB.92.081201}
\begin{equation}
\gamma=\oint_\mathcal{C} \mathcal{A}(\mathbf{k})\cdot d\mathbf{k},
\end{equation}
where $\mathcal{A}(\mathbf{k})$ is the Berry connection, and $\mathcal{C}$ is a closed loop in three-dimensional momentum space. The corresponding Berry curvature is $\mathbf{\Omega}(\mathbf{k})=\nabla \times \mathcal{A}(\mathbf{k})$. As shown in Fig. \ref{FIG3}(c), we calculate the Berry curvature $\mathbf{\Omega}(\mathbf{k})$ along an arbitrary line $W_1$-$\Gamma$-$W_2$ in the $k_y$-$k_z$ plane. This figure shows that $\mathrm{\Omega}_{x}(\mathbf{k})$ has opposite peaks near the momentum positions of crossing points along the opposite directions, which is in accordance with the coexistence of $\mathcal{I}$ and $\mathcal{T}$ symmetries. To further reveal the nontrivial type-II Dirac nodal lines in $\mathrm{CsV_6Sb_6}$, we calculate a variation of Berry phase $\gamma$ in the $k_y$-$k_z$ plane, which corresponds to the one-dimensional system along the $W_1$-$\Gamma$-$W_2$. As shown in Fig. \ref{FIG3}(d), there is a jump of $\pi$ ($-\pi$) when the closed path $\mathcal{C}$ encircles the nodal lines from $W_1$ ($W_2$) to $\Gamma$, indicating its nontrivial topology.

\begin{figure}
    \centering
    \includegraphics[scale=0.183]{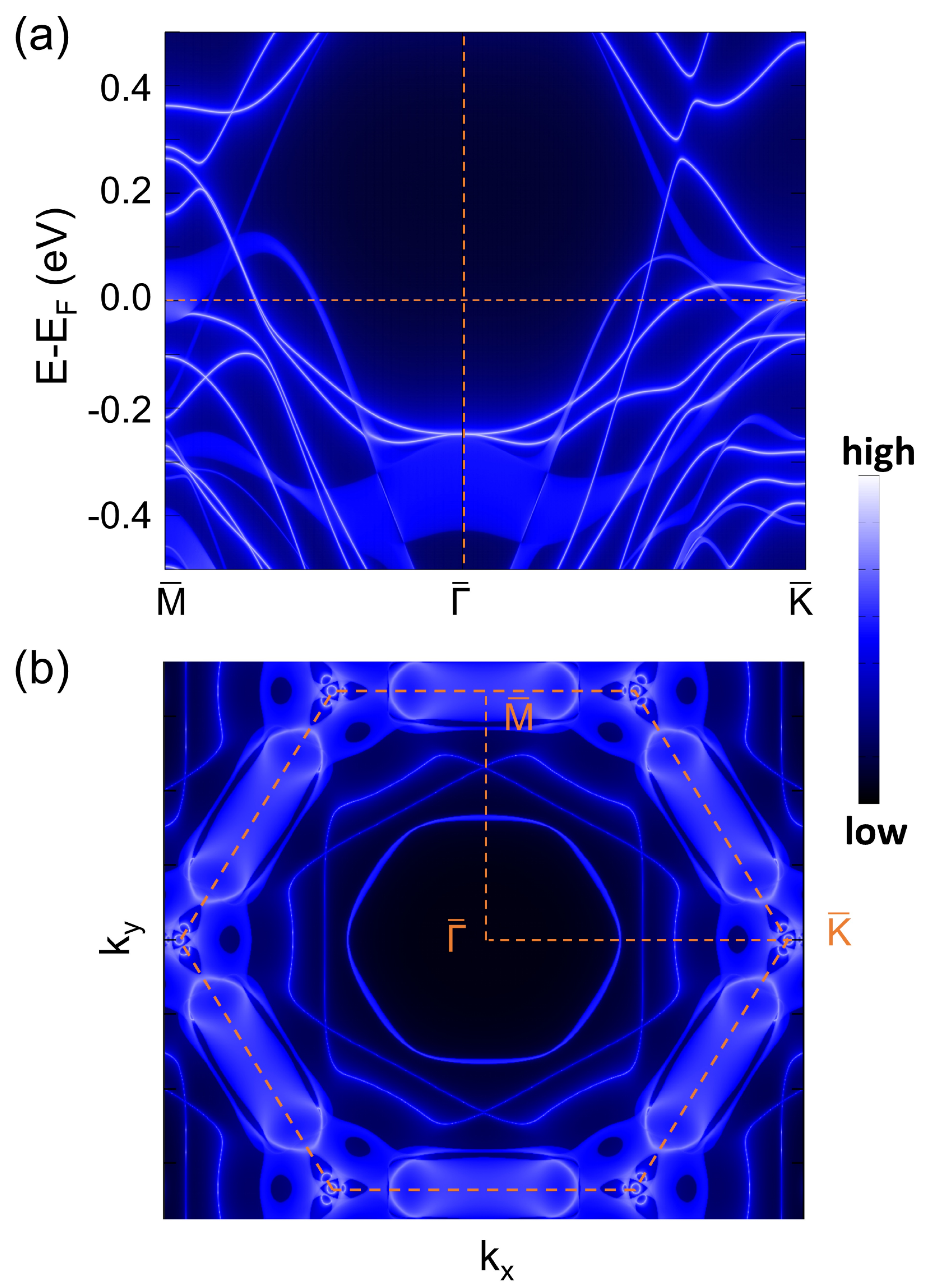}
    \caption{Surface states of $\mathrm{CsV_6Sb_6}$. (a) The calculated LDOS and (b) Fermi surface projected on the semi-infinite (111) surface of $\mathrm{CsV_6Sb_6}$. In panel (b), there are two Fermi arcs well separated from the bulk bands enclosed to the $\bar{\Gamma}$ point. The high-symmetry lines and the first BZ of (111) surface are marked by yellow-dashed lines.
    \label{FIG4}}
\end{figure}

For the nodal lines protected by $\mathcal{I}$ and $\mathcal{T}$ symmetries, the spin-orbital coupling (SOC) can always gap out the nodal lines and drive the system to other topological phases, such as topological insulators, Dirac semimetals, and even topologically trivial phases \cite{PhysRevLett.115.036806, C7NR03520A}. Indeed, the calculated band structures within SOC of $\mathrm{CsV_6Sb_6}$ indicate that two crossing bands without SOC now belong to the same mirror eigenvalues $\Gamma_4$ of $C_s$ and thus the band crossings are avoided (see Fig. S2 in the SM \cite{SM}). However, the rather weak SOC effect of $\mathrm{CsV_6Sb_6}$ only leads to a band gap width less than $\sim$1 meV. That is to say, its type-II Dirac nodal line fermions are nearly intact.  The nodal line phases usually have the same $\mathbb{Z}_2$ invariant as its gapped topological phases \cite{PhysRevLett.115.036806}. To verify the topological classification of $\mathrm{CsV_6Sb_6}$, we calculate the parity eigenvalues of occupied bands at the time-reversal invariant momenta points (see the details in the SM \cite{SM}). The results reveal that the band gaps opened by SOC in $\mathrm{CsV_6Sb_6}$ are topologically nontrivial with a (0;111) weak $\mathbb{Z}_2$ index. Moreover, we also find that other nontrivial bands are present near the Fermi level \cite{SM}, exhibiting entangled topological features. The type-II Dirac nodal line fermions categorized as a $\mathbb{Z}_2$ classification are expected to give rise to exotic quantum transport phenomena.

The type-II Dirac nodal lines combining with $\mathbb{Z}_2$-type nontrivial band topology in close proximity to the Fermi level can give rise to topological surface states. To illustrate this, we calculate the surface states of $\mathrm{CsV_6Sb_6}$ using the iterative Green's function method based on Wannier TB Hamiltonian as implemented in the WANNIERTOOLS package \cite{Sancho1984, WU2017}. The calculated local density of states (LDOS) and Fermi surface projected on the semi-infinite (111) surface of $\mathrm{CsV_6Sb_6}$ are shown in Fig. \ref{FIG4}. As depicted in Fig. \ref{FIG4}(a), we can see that the projected type-II Dirac cones  along $\bar{\Gamma}$-$\bar{M}$ exhibit a band width due to the tiny energy dispersion of nodal-line. The surface states along $\bar{\Gamma}$-$\bar{M}$ and $\bar{\Gamma}$-$\bar{K}$ exhibit apparent anisotropy but are degenerate at the $\bar{\Gamma}$ point, indicating $\mathbb{Z}_2$-type band topology of $\mathrm{CsV_6Sb_6}$. As shown in Fig. \ref{FIG4}(b), the Fermi surface projected to (111) surface shows that there are two closed nontrivial Fermi arcs crossing along $\bar{\Gamma}$-$\bar{M}$. This is in accordance with Dirac nodal lines lying in the mirror planes. Importantly, the nontrivial Fermi arcs are well separated from the bulk projected bands, facilitating their detection in experiments and further exploration.

In conclusion, we have studied the electronic and topological properties of a latest class of experimentally synthesized layered compounds \textit{A}$\mathrm{V_6Sb_6}$ with a kagome $(\mathrm{V_3Sb})_2$ bilayer \cite{A166arXiv}. Based on first-principles calculations and topological analysis, we reveal that these novel \textit{A}$\mathrm{V_6Sb_6}$ compounds possess type-II Dirac nodal line fermions close to the fermi level and exhibit $\mathbb{Z}_2$-type nontrivial band topology. Importantly, it is worth noting that \textit{A}$\mathrm{V_6Sb_6}$ as well as \textit{A}$\mathrm{V_3Sb_5}$ can be represented by a generic chemical formula \textit{A}$_{m-1}$Sb$_{2m}$(V$_3$Sb)$_n$. Considering that superconductivity in \textit{A}$\mathrm{V_6Sb_6}$ was realized under pressure \cite{A166arXiv}, our findings are expected to extend the knowledge toward understanding quantum ordered states and unconventional topological fermions beyond standard model in the \textit{A}-V-Sb kagome family.

~~~\\

This work is supported by the National Natural Science Foundation of China (11888101 and 11974062), the National Key Research and Development Program of the Ministry of Science and Technology of China (2017YFA0303001, 2019YFA0704901, and 2016YFA0300201), the Anhui Initiative in Quantum Information Technologies (AHY160000), the Strategic Priority Research Program of the Chinese Academy of Sciences (XDB25000000), the Science Challenge Project of China (TZ2016004), and the Key Research Program of Frontier Sciences, CAS, China (QYZDYSSWSLH021). The DFT calculations in this work are supported by the Supercomputing Center of University of Science and Technology of China.

%\bibliographystyle{apsrev4-2}
%\bibliography{ref}

\newpage
\begin{widetext}

\setcounter{figure}{0}
\setcounter{equation}{0}
\makeatletter

\makeatother
\renewcommand{\thefigure}{S\arabic{figure}}
\renewcommand{\thetable}{S\Roman{table}}
\renewcommand{\theequation}{S\arabic{equation}}

\begin{center}
	\textbf{
		\large{Supplemental Material for}}
	\vspace{0.2cm}
	
	\textbf{
		\large{
			``Type-II Nodal Line Fermions in New $\mathbb{Z}_2$ Topological Metals \textit{A}$\mathbf{V_6Sb_6}$ (\textit{A}=K, Rb, and Cs) with  Kagome Bilayer" }
	}
\end{center}

In this  Supplemental Materials, we present lattice parameters of \textit{A}$\mathrm{V_6Sb_6}$ (\textit{A}=K, Rb, and Cs) compounds from first-principles calculations and experiments, electronic band structures of of $\mathrm{RbV_6Sb_6}$  and $\mathrm{KV_6Sb_6}$ in the absence of spin-orbital coupling (SOC), electronic band structures of $\mathrm{CsV_6Sb_6}$ in the presence of SOC, and parity analysis of $\mathrm{CsV_6Sb_6}$ at the time-reversal invariant momenta points.

~~~\\
~~~\\

The \textit{A}$\mathrm{V_6Sb_6}$ (\textit{A}=K, Rb, and Cs) compounds crystallize in a rhombohedral structure with space group $R\bar{3}m$ (No. 166). The primitive lattice vectors of the rhombohedral representation can be expressed as
\begin{equation}
\begin{split}
&\mathbf{t}_1=-\frac{a}{2}\mathbf{i} -\frac{\sqrt{3}}{6}a \mathbf{j}+ \frac{c}{3}\mathbf{k},\\
&\mathbf{t}_2=\frac{a}{2}\mathbf{i} -\frac{\sqrt{3}}{6}a \mathbf{j}+ \frac{c}{3}\mathbf{k},\\
&\mathbf{t}_3= \frac{\sqrt{3}}{3}a \mathbf{j}+ \frac{c}{3}\mathbf{k},
\end{split}
\end{equation}
where $a$ and $c$ are lattice constants of the hexagonal representation as included in Table \ref{S1}.

\begin{table}[H]
  \centering
\caption{Lattice parameters of \textit{A}$\mathrm{V_6Sb_6}$ (\textit{A}=K, Rb, and Cs) compounds from first-principles calculations and experiments. $a$ is the
lattice constant in the kagome (or hexagonal) plane and $c$ is the lattice constant along its stacking direction. The comparison shows a good agreement.
\label{S1}}
\setlength{\tabcolsep}{8mm}{
\begin{tabular}{c|cccc}
  \hline
  \hline
    &   & $\mathrm{CsV_6Sb_6}$  & $\mathrm{RbV_6Sb_6}$  & $\mathrm{KV_6Sb_6}$  \\
  \hline
  Space group & \multicolumn{4}{c} {$R\bar{3}m$ (No. 166)} \\
  \hline
  $a$ ({\AA}) & this work & 5.43 & 5.42 & 5.42 \\
      & expt. & 5.51 & 5.51 &  / \\
  $c$ ({\AA}) & this work & 34.94 & 34.23 &33.69 \\
      & expt. & 35.28 & 34.61 & /  \\
  \hline
  \hline
\end{tabular}}
\end{table}

\begin{figure}[H]
    \centering
    \includegraphics[scale=1.5]{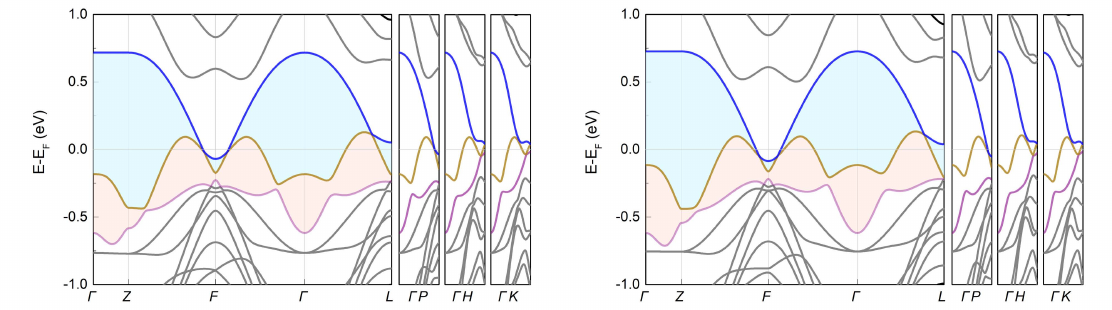}
    \caption{Electronic band structures of $\mathrm{RbV_6Sb_6}$ (left panel) and $\mathrm{KV_6Sb_6}$ (right panel) along high-symmetry lines in the absence of SOC. Three bands 115, 117, 119 are colored by pink, dark-yellow, and blue, respectively.
    \label{FIGS1}}
\end{figure}

\begin{figure}[H]
    \centering
    \includegraphics[scale=1.4]{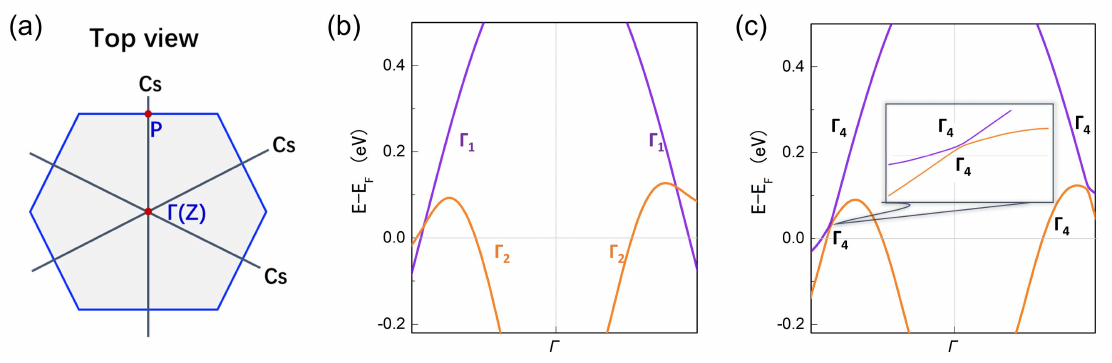}
    \caption{(a) The top views of a rhombohedral Brillouin zone (BZ). There are three high-symmetry middle planes of the BZ with little group $C_s$, which are with respect to three-fold rotational symmetry $C_{3z}$. (b) The enlarged view of bands of $\mathrm{CsV_6Sb_6}$ in the absence of spin-orbital coupling (SOC) along an arbitrary line in a middle plane. Two crossing bands belong to opposite mirror eigenvalues $\Gamma_1$ (+1) and $\Gamma_2$ (-1) of $C_s$. (c) In the presence of SOC, two crossing bands without SOC now belong to the same mirror eigenvalues $\Gamma_4$ of $C_s$ and thus the band crossings are avoided.
    \label{FIGS2}}
\end{figure}

\begin{table}[H]
  \centering
  \caption{The irreducible representations (irrep.) and parity products (prod.) of occupied bands of $\mathrm{CsV_6Sb_6}$ at the time-reversal invariant momenta points. The band topology between the lowest conduction band (119) and highest valence band (117) characterized by the (0;111) $\mathbb{Z}_2$ index. The nontrivial band topology are also present when the occupied band is band 115 or band 113, which is close to the Fermi level.
\label{S2}}
  \setlength{\tabcolsep}{4mm}{
  \begin{tabular}{c|cccc|cccc|c}
    \hline
    \hline
    Occ. band &\multicolumn{4}{c}{irrep.} & \multicolumn{4}{c}{Parity Prod.} & Invar. \\
    \hline
      & $\Gamma$  & $3\times F$ & $3\times L$ & Z        &{$\delta_{\Gamma}$} & $\delta_{F}$ & $\delta_{L}$ & $\delta_{Z}$ & $\mathbb{Z}_2$ \\
      & ($D3d$)   &$(C2h)$     &$(C2h)$       & ($D3d$)  &                    &              &              &              &                \\
    \hline
    119 & $\Gamma_{4}^{-}$ & $\Gamma_{3}^{+}\oplus\Gamma_{4}^{+}$              &$\Gamma_{3}^{-}\oplus\Gamma_{4}^{-}$ & $\Gamma_{4}^{+}$ & - & - &              - & - & (0;000) \\
    117 & $\Gamma_{4}^{+}$ & $\Gamma_{3}^{+}\oplus\Gamma_{4}^{+}$              &$\Gamma_{3}^{-}\oplus\Gamma_{4}^{-}$ & $\Gamma_{4}^{-}$ & + & - &              + & - & (0:111) \\
    115 & $\Gamma_{4}^{-}$ & $\Gamma_{3}^{-}\oplus\Gamma_{4}^{-}$              &$\Gamma_{3}^{+}\oplus\Gamma_{4}^{+}$ & $\Gamma_{4}^{+}$ & + & - &              - & + & (0;111) \\
    113 & $\Gamma_{5}^{-}\oplus \Gamma_{6}^{-}$ & $\Gamma_{3}^{-}\oplus\Gamma_{4}^{-}$ &$\Gamma_{3}^{+}\oplus\Gamma_{4}^{+}$ & $\Gamma_{5}^{+}\oplus\Gamma_{6}^{+}$ & - & + & - & + & (0;111) \\
    \hline
    \hline
  \end{tabular}}
\end{table}

\end{widetext}
\end{document}